# The anomalies of the properties of nanomaterials related to the distribution of the grain sizes

M.D.Glinchuk[1], P.I.Bykov[2]

[1]Institute for Problems of Materials Science, NASc of Ukraine, Kjijanovskogo 3, 03680 Kiev 142, Ukraine, dep4@materials.kiev.ua

[2]Radiophysics Faculty of Taras Shevchenko National University of Kiev, 2, Acad. Glushkov Ave., building # 5, 03127 Kiev, Ukraine

We have performed the calculations of the size effect in the temperature dependence of $BaTiO_3$ nanograin ceramics specific heat and dielectric permittivity. We took into account the distribution of the grain sizes, that exists in any real nanomaterial. This distribution lead to the distribution of the temperatures of the size driven transition from ferroelectric to paraelectric phase because of relation between the temperature and the sizes. We calculated the transition temperature distribution function on the basis of the sizes distribution function. This function allowed to calculate the temperature dependence of any physical quantity in a nanomaterial. As an examples we calculated specific heat and dielectric permittivity in nanograin ferroelectric ceramics. The results demonstrate the strong influence of the size distribution on the observed properties and especially on extracted from experiment values of critical size and temperature. We carried out the comparison of the theory with the measured specific heat and dielectric permittivity in $BaTiO_3$ nanomaterial. The developed theory described the experimental data pretty good. The possibility of the extraction of size distribution function parameters as well as real values of critical parameters from experimental data is discussed.

## 1. Introduction

The anomalies of physical properties of nanomaterials, namely nanoparticle powders and nanograin ceramics attract the growing interest of scientists and engineers because of size effects of properties useful for applications [1,2,3]. In ferroelectric nanomaterials the most important size effect is known to be the transformation of ferroelectric phase into paraelectric one at some critical size [4]. Investigation of this phenomenon was performed experimentally and theoretically in several works (see e.g. [5,6,7]). However the most of these works were devoted to investigation of dielectric properties. The experimental study of the thermal properties e.g. specific heat in $BaTiO_3$ polycrystalline thin film and nanograin thick film was published only recently [8,9,10]. Two main effects were revealed, namely the temperature of specific heat jump appeared to be dependent on average film thickness or nanoparticle size, and there were the distributions of these temperatures, its width became larger with the temperature decrease. The position of the maximum of this distribution was reasonably supposed to be related to the temperature of size driven ferroelectric–paraelectric phase transition. The empirical expression for the transition temperature dependence on an average particle size was derived from experimental points. The physical mechanisms, which lead to this expression and to the distribution of the transition temperatures, were not discussed in [8,9,10]. The measurements of dielectric permittivity dependence on average grain size in $BaTiO_3$ nanograin ceramics lead to the "puzzle" of much larger (about 10 times) value of critical size in the ceramics than in nanopowder [11]. Up to now the physical reasons of this large difference stayed unclear.

In present paper we described the main experimental results about size effects of specific heat and dielectric susceptibility in nanograin $BaTiO_3$ on the basis of the equations obtained by us earlier [5]. We took into account also the distribution of the particle sizes, that really exists in any nanomaterial. We had shown that this distribution leads to the distributions of the transition temperature. It was shown that all observed properties have to be smeared and their maxima positions have to be shifted by these distributions. The developed theory described the observed in nanograin $BaTiO_3$ ceramics specific heat temperature dependence and dielectric permittivity size dependence pretty good.

## 2. The theoretical description of the specific heat in nanomaterials

The calculation of nanomaterials properties used to be performed in phenomenological theory framework on the basis of free energy functional variation (see e.g. [6]). This procedure leads to differential Euler–Lagrange equation with the boundary conditions originated from surface energy. In the majority of the papers the solution of the equation and the calculations of some dielectric properties were performed numerically. The method of analytical calculation was proposed recently [5] (see also [12,13]). It was shown that the properties can be obtained by minimization of conventional type of free energy, but with coefficient before square polarization, that depends on particle size, temperature, contribution of depolarization field and extrapolation length.

This free energy view is the following [5]:

$$F = \frac{A_R}{2} P^2 + \frac{B_R}{4} P^4 + \frac{C_R}{6} P^6 - P \cdot E \quad (1)$$

Here $P$ is averaged over nanopartical volume polarization, $E$ is external electric field, $B_R \approx b$, $C_R \approx c$, where $b$ and $c$ are the corresponding constants of bulk material. The renormalized coefficient $A_R$ has the form:

$$A_R \approx \alpha \cdot (T - T_{cl}(R)) \quad (2)$$

where $\alpha$ is inverse Curie-Weiss constant of the bulk, $R$ is the size of the spherical nanoparticles. The temperature of size driven phase transition $T_{cl}$ can be approximately written as

$$T_{cl}(R) \approx T_c \left(1 - \frac{R_{cr}(0)}{R}\right) \quad (3)$$

$$R_{cr}(T) \approx \frac{R_{cr}(0)}{1 - \frac{T}{T_c}} \quad (4)$$

Here $T_{cl}(R)$ and $R_{cr}(T)$ are critical temperature and radius of the phase transition at some arbitrary radius $R$ or temperature $T$ respectively, $T_c$ is phase transition temperature of bulk material. Substitution of Eqs. (3), (4) into Eq. (2) transforms $A_R$ into

$$A_R \approx \alpha (T - T_c)\left(1 - \frac{R_{cr}(T)}{R}\right) \quad (5)$$

The Eqs. (2)–(5) allow to calculate temperature and size dependence of all nanomaterial properties averaged over the particle volume by conventional minimization of free energy (1). For example, dielectric permittivity has the form

$$\varepsilon_{PE}(T,R) = \begin{cases} \dfrac{\varepsilon_0}{(R_{cr}(T)/R - 1)}, & R < R_{cr} \\ \dfrac{1}{\alpha(T - T_{cl})}, & T > T_{cl} \end{cases} \quad (6a)$$

$$\varepsilon_{FE}(T,R) = \begin{cases} \dfrac{\varepsilon_0}{2(1 - R_{cr}(T)/R)}, & R > R_{cr} \\ \dfrac{1}{2\alpha(T_{cl} - T)}, & T < T_{cl} \end{cases} \quad (6b)$$

$$\varepsilon_0 = \frac{1}{\alpha(T_c - T)}$$

where $\varepsilon_{PE}$ and $\varepsilon_{FE}$ are respectively permittivity in paraelectric and ferroelectric phase and the first or second lines in brackets can be used respectively at some fixed temperature or radius. Keeping in mind that we are interested in consideration of the thermal capacity $C_p$ in BaTiO$_3$ let us write $C_p$ for the phase transition of the I$^{st}$ order on the basis of Eqs. (1), (2). Allowing for $C_p = -T \dfrac{d^2\Phi}{dT^2}$, one obtains the difference $C_p(T < T_{cl}) - C_p(T > T_{cl}) \equiv \Delta C_p$ in the form:

$$\Delta C_p = \dfrac{\alpha^2}{2b} \dfrac{T}{\sqrt{1 + \dfrac{4\alpha c}{b^2}(T_{cl}(R) - T)}}, \qquad T < T_{cl} \tag{7}$$

Note, that for the phase transition of the I$^{st}$ order the transition temperature $T_{cl}$ written for the phase transition of the II$^{nd}$ order in the form of Eq. (3) has to be shifted on the value $\Delta T = \dfrac{3}{16}\dfrac{b^2}{\alpha c}$ [5].

### 3. Distribution function of transition temperature

In real nanomaterials the sizes of nanoparticles are usually distributed, the form and parameters of the distribution function being dependent on the technology of a sample preparation. Let us suppose, that the distribution function of radius $R$ has Gaussian form, namely

$$f(R) = C \exp\left(-\left(\dfrac{R - R_0}{\sigma}\right)^2\right), \qquad 0 \le R \le \infty \tag{8a}$$

where $C$ is normalization constant

$$C = \dfrac{2}{\sigma\sqrt{\pi}\left(erf\left(\dfrac{R_0}{\sigma}\right) + 1\right)}, \tag{8b}$$

In Eqs.(8) $R_0$ and $\sqrt{\ln 2}\,\sigma$ are respectively the most probable radius and half-width on half-height. Allowing for in many experimental works average radius $\overline{R}$ of nanoparticles (obtained e.g. on the basis of X-ray diffraction method) is given, it is useful to write the relation between $\overline{R}$ and $R_0$:

$$\overline{R} = R_0 + \dfrac{\sigma \exp\left(-\left(\dfrac{R_0}{\sigma}\right)^2\right)}{\sqrt{\pi}\left(1 + erf\left(\dfrac{R_0}{\sigma}\right)\right)}, \tag{9}$$

In Fig.1 one can see, that $\overline{R} \approx R_0$ at $\dfrac{R_0}{\sigma} \ge 1{,}5$, while at smaller value there is a difference between them. In particular at $R_0 \to 0$ the value $\dfrac{\overline{R} - R_0}{\sigma} \to \dfrac{1}{\sqrt{\pi}}$.

It follows from Eq. (3) that the distribution of radius has to be the source of the distribution of transition temperatures $T_{cl}$. In accordance with the theory of probability [14] the distribution function $F(T_{cl})$ can be expressed via $f(R)$ by the following way:

$$F(T_{cl}) = f(R)\left|\dfrac{dR}{dT_{cl}}\right|, \tag{10}$$

Eq. (10) with respect to Eqs. (3), (8) yields:

$$F(T_{cl}) = C_1 \dfrac{R_{cr}(0)T_c}{(T_c - T_{cl})^2} \exp\left(-\dfrac{R_{cr}^2(0)T_c^2}{\sigma^2}\left(\dfrac{1}{T_c - T_{cl}} - \dfrac{1}{T_c - T_{cl}^0}\right)^2\right), \tag{11a}$$

$$C_1 = \dfrac{2}{\sigma\sqrt{\pi}\left(1 + erf\dfrac{R_{cr}(0)T_c}{\sigma(T_c - T_{cl}^0)}\right)}, \tag{11b}$$

where $T_{cl}^0 \equiv T_{cl}(R = R_0)$ is the most probable transition temperature.

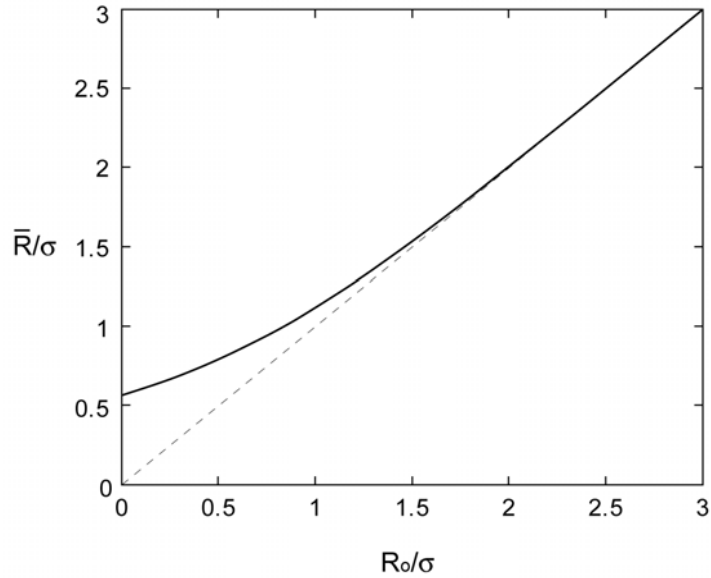

Figure 1. The dependence of average grain size $\overline{R}$ on the most probable grain size $R_0$ and dispersion parameter σ.

Using the distribution functions in the form of Eqs. (8) or Eqs. (11) one can average any physical property written as a function of the nanoparticles radius or temperature. For example, when calculating the dielectric permittivity, it is possible to average with $f(R)$ or $F(T_{cl})$ the expression in the first or in the second lines respectively of Eqs. (6). The results of the averaging are depicted in Figs. 2, 3.

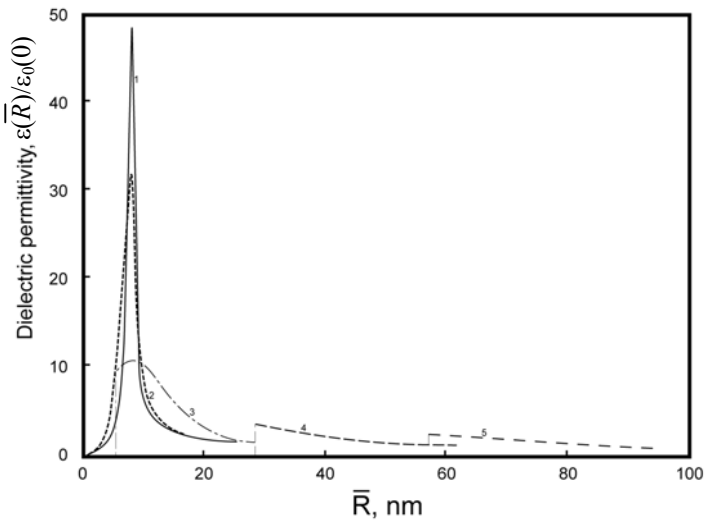

Figure 2. The dependence of the relative dielectric permittivity on the average grain size $\overline{R}$ calculated on the basis of Eqs. (12a), (8) for the different dispersion parameters σ: 1 (1), 2 (2), 10 (3), 50 (4), 100 (5).

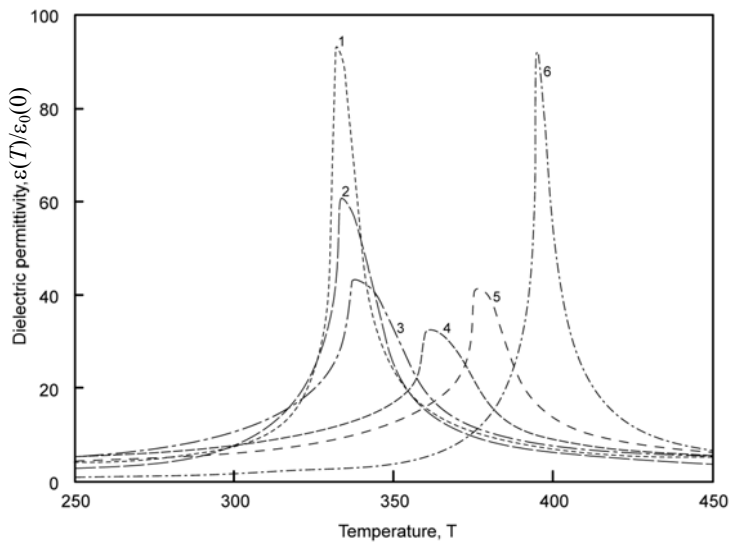

Figure 3. The dependence of the relative dielectric permittivity on the temperature, calculated on the basis of Eqs. (12b), (11) for the following values of parameter $R_0/σ$: 26 (1); 5.2 (2), 2.6 (3), 0.65 (4), 0.26 (5), $5.2 \cdot 10^{-3}$ (6).

Figs. 2 and 3 are built respectively on the basis of the following Equations:

$$\varepsilon(R_0, T) = \frac{\varepsilon_0(0)}{q|1-t|} \int_0^\infty \frac{f(R)dR}{\sqrt{\left(1 - \frac{R_{cr}(T)}{R}\right)^2 + \delta^2}} \qquad (12a)$$

and

$$\varepsilon(t) = \frac{\varepsilon_0(0)}{q} \int_0^1 \frac{F(x)dx}{\sqrt{(t-x)^2 + \delta^2}}, \qquad (12b)$$

where $x \equiv \frac{T_{cl}}{T_c}$, $t \equiv \frac{T}{T_c}$ and $\delta = 0{,}01$ is small parameter introduced to make maximum height of permittivity restricted, $q = 2$ or $1$ in ferroelectric or paraelectric phase respectively, $\varepsilon_0(0) = \frac{1}{\alpha T_c}$.

More detailed discussion of the influence of the sizes distribution on dielectric susceptibility and the peculiarities depicted in Figs. 2, 3 we will discuss later.

### 4. Comparison of calculated and measured specific heat

For the sake of theoretical description of observed in nanograin BaTiO$_3$ ceramics temperature dependence of specific heat we performed the averaging of Eq. (7) with the help of distribution function $F(T_{cl})$ in the form of Eqs. (11). Namely we carried out the calculation of the integral

$$\overline{\Delta C_p}(T_{cl}^0, T, \sigma) = \frac{\alpha^2 T}{2b} \int_0^{T_c} \frac{F(T_{cl}, T_{cl}^0, \sigma)dT_{cl}}{\sqrt{1 + \frac{4\alpha c}{b^2}(T_{cl} - T)}}, \qquad (13a)$$

Allowing for the relation between half-width on half-height of $f(R)$ and $F(T_{cl})$ ($T_{h.w.}$), namely

$$\sigma\sqrt{\ln 2} = \frac{T_c \Delta T_{h.w.} R_{cr}(0)}{(T_c - T_{cl}^0)(T_c - T_{h.w.})}, \quad \Delta T_{h.w.} = T_{h.w.} - T_{cl}^0 \qquad (13b)$$

it appeared possible to extract $\sigma$ from the observed $\Delta T_{h.w.}$ values and then $R_0$ values from Fig. 1. The obtained data and experimental parameters are presented in the Table 1. They illustrate the possibility of extraction of the parameters $\sigma$ and $R_0$ of the distribution function from experimental data. To obtain $\sigma$ and $R_0$ values given in Table 1 we took $R_{cr}(0) = 4$ nm [10] and the values of $T_{cl}^0$ were calculated on the basis of Eq. (3) at $R = R_0$.

Table 1. The experimental data and the parameters of the grain sizes distribution function extracted from observed temperature dependence of BaTiO$_3$ nanograin ceramic specific heat.

| Experiment [10] | $\overline{R}$, nm | 82,5 | 45,0 | 32,5 | 17,5 |
|---|---|---|---|---|---|
| | $T_m$, K | 393,0 | 385,8 | 372,0 | 332,5 |
| | $\Delta T_{h.w.}$, K | 0,3 | 5,2 | 7,8 | 8,1 |
| Theory | $R_0$, nm | 82,5 | 45,0 | 32,5 | 17,5 |
| | $\sigma$, nm | 4,000 | 7,168 | 5,302 | 1,807 |

In Fig. 4 one can see the comparison of the calculated $T_{cl}^0 = T_{cl\,max} \equiv T_m$ with experimental data obtained for several $\overline{R}$ sizes, allowing for the considered case of small enough half-width the value of $R_0$ practically coincides with $\overline{R}_{max}$. One can see from Fig. 4 that the theory fits experimental points very good. It should be noted that although the measurements were performed on 500 nm BaTiO$_3$ film with different grain sizes, in the films with the thickness more than 400 nm the specific heat practically coincides with that in bulk (see [8,9]), so that the 500 nm $BaTiO_3$ film can be considered as the bulk ceramic.

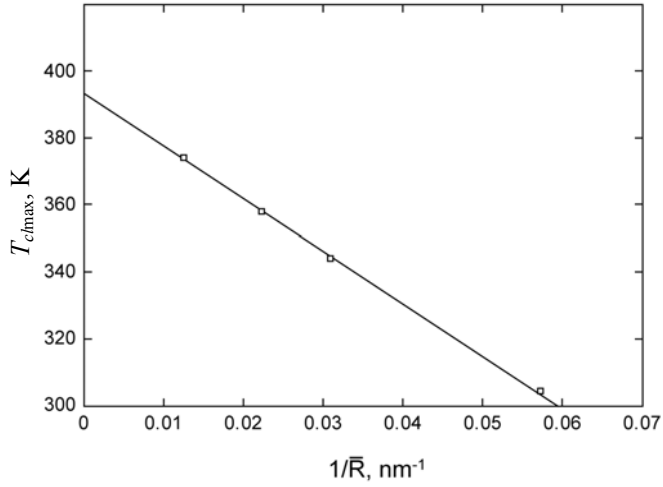

Figure 4. The dependence of the ferroelectric–paraelectric phase transition temperature on the inverse average grain size for BaTiO$_3$ nanograin ceramic. Solid line – theory, squares – experiment [10].

Therefore, the only parameter that was taken from measurements of the specific heat is the value of half-width $\Delta T_{h.w.}$ because $T_{cl}^0$ can be calculated via $R_0$. Keeping in mind the I$^{st}$ order phase transitions in BaTiO$_3$, we calculated the shift of transition temperature as $\Delta T_{cl}^0 = \frac{3}{16}\frac{b^2}{\alpha c} \approx 28\,°C$ for the values of parameters taken from [15] for BaTiO$_3$ bulk material. The results of the theoretical calculations on the basis of Eqs. (13) with respect to Eqs. (11) and the values of σ given in Table 1 are depicted in Fig. 5 by solid line. One can see that this line fits the experimental points pretty good. Note, that the slope of the curves is related to thermal capacity in paraelectric phase ($T > 400\,°C$).

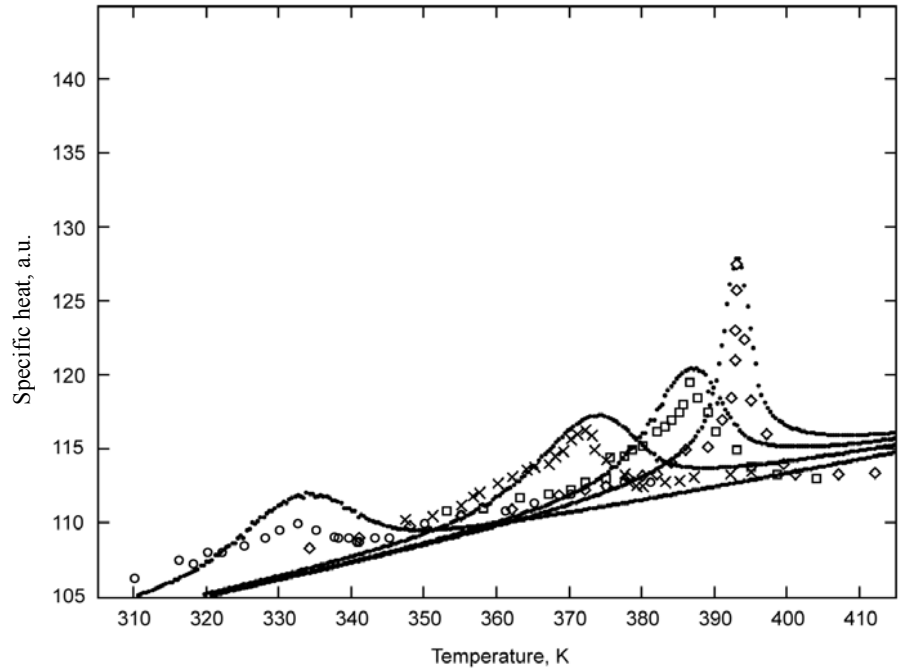

Figure 5. Temperature dependence of the specific heat of 500 nm BaTiO$_3$ films with the different grain size, calculated on the basis of Eqs. (11), (13) (solid line) and experimental data taken from [10] for the following grain sizes: 35 nm (○), 65 nm (×), 90 nm (□), 165 nm (◊). The values of the experimental data and fitting parameters are given in Table1.

**5. Discussion**
5.1. Thin films roughness as possible source of transition parameters distribution
Experimental data obtained in [8,9] for temperature dependences of BaTiO$_3$ specific heat for the films with different thickness look like those for ceramics with nanosizes of the grains (see Fig. 5). In particular it was shown that when the thickness of the film reduces, the phase transition temperature decreases while a smearing of the anomaly increases. The anomaly is quite weak for 40 nm film and it was not detected for 20 nm film. Authors draw the attention to the fact that the sharp increase of a film roughness was revealed for the ultrathin films. To our mind the latter can be the reason of diffusivity of the specific heat anomaly near thickness induced phase transition from ferroelectric to paraelectric phase in the thin films [8,9]. Therefore, the distribution of the film thickness has to be taken into account when considering thin films properties. In particular the calculations of specific heat in the films can be performed similarly to the calculations in the section 4 for nanograin ceramics, allowing for the temperature of thickness induced phase transition can be written in the following form [12]:

$$T_{cl} = T_c \left[ 1 - \frac{l_0^2(0)}{l} \left( \frac{1}{\lambda_1 + l_d} + \frac{1}{\lambda_2 + l_d} \right) \right], \quad (14)$$

Here $l_0^2(0) = \frac{\gamma}{\alpha_0 T_c}$, $l_d^2 = \frac{\gamma}{4\pi}$, and $l$, $\lambda_1$, $\lambda_2$, $\gamma$ and $\alpha_0$ are respectively a film thickness, extrapolation lengths, coefficient before squared polarization gradient in free energy functional and inverse Curie–Weiss constant. Comparison of Eqs. (14) and (3) shows, that the dependence of $T_{cl}$ on particle sizes $R$ or film thickness $l$ is of the same type. The detailed calculations of specific heat anomalies in thin films allowing for the difference in geometry of the films and nanoparticles is in a progress now. The comparison of the calculated and observed anomalies will give valuable information about the parameters of the film thickness distribution function.

5.2. Influence of sizes distribution function on critical temperature and radius.

It is generally believed, that critical temperature and radius of size driven ferroelectric-paraelectric phase transition can be obtained from the peculiar points of the properties, e.g. from maximum of dielectric permittivity. However in real materials there is a distribution of transition temperature $T_{cl}$ (see Eq. (11a)) related to the distribution of sizes. In general case the physical reason of the uncertainty of physical meaning of the parameters, which correspond to the observed property maxima, is the competition between distribution function and the property maxima positions. In particular for dielectric permittivity it is at $R = R_0$ and $R = R_{cr}$ (see Eqs. (8a) and (12a) respectively). So that the distribution of the particles sizes makes it unclear whether the position of observed $\varepsilon(R)$ maximum obligatory coincides with $R_{cr}$ value. Let us consider this in more the details.

First of all when the width of distribution function is very small ($\sigma \to 0$), i.e. it can be represented as δ-function, there is one $T_{cl}$ value only and so the position of $\varepsilon(T)$ or $\varepsilon(\overline{R})$ maxima indeed define this critical temperature or critical radius respectively. This statement can be correct for small enough $\sigma \neq 0$ also. This situation, that depends on the samples preparation technology, took place in BaTiO$_3$ nanograin materials investigated in [8,9,10]. The latter is related to the quantitative criterion $\overline{R} \approx R_0$ at $R_0/\sigma \geq 1,5$ (see Table 1). It follows from Eq. (9), that at $\overline{R} \approx R_0$ the contribution of the second term can be neglected similarly to the limit $\sigma \to 0$. Therefore the criterion $\frac{\overline{R}}{\sigma} \approx \frac{R_0}{\sigma} \geq 1,5$ can be considered as the condition for extracting of critical parameters (temperature and radius) from the properties maxima positions, i.e. one can write the necessary relation between the distribution function parameters as

$$\sigma < \frac{2}{3} R_0 \quad (15a)$$

But in many real samples this criterion is not fulfilled. When $\frac{R_0}{\sigma} < 1,5$ or $\frac{\overline{R}}{\sigma} < 1,5$ the difference $\overline{R} - R_0$ increases with σ increase. Even at some finite value of σ, but at $R_0 \to 0$ average radius $\overline{R} \to \frac{\sigma}{\sqrt{\pi}}$ so that $\overline{R}$ is restricted by this value, i.e. $\overline{R}_{min} = \frac{\sigma}{\sqrt{\pi}}$ (see Fig. 1). The same limit can be achieved at $R_0 \neq 0$, $\sigma \to \infty$, that gives $\overline{R}_{min} \to \infty$. The latter case correspond to bulk materials, while the former case shows that in the nanomaterials there is the restriction of $\overline{R}$ related to σ values. It is obvious that with σ increase $\overline{R}_{min}$ can become larger than $R_{cr}$ value so that it can be impossible to extract $R_{cr}$ from experimental data. To obtain real value of $R_{cr}$ the sizes distribution function width has to satisfy the condition

$$\frac{\sigma}{\sqrt{\pi}} < R_{cr}(T) \qquad (15b)$$

To illustrate this we depicted in Fig. 2 dielectric permittivity $\varepsilon(\overline{R})$ dependence for $R_{cr}(T = 196$ K$) = 8$ nm [10]. One can see, that the condition (15b) is fulfilled for the curves 1, 2, 3 and so their maxima positions correspond to $\overline{R}_{max} \approx R_{cr}(T)$. The curves 4 and 5 are strongly shifted from $R_{cr}$ value because for them $\overline{R}_{min} = \frac{\sigma}{\sqrt{\pi}}$ is about 25 nm and 50 nm respectively which are several times larger than $R_{cr}(T) = 8$ nm.

It should be noted, that conditions (15a) and (15b) coincide with one another at $R_0 = \frac{3}{2}\sqrt{\pi}R_{cr}$, i.e. the $R_0$ has to be about 3 times larger, than $R_{cr}$. Separately, Eq. (15a) reflects the desirable quality of nanograin ceramics, while Eq. (15b) is the necessary condition for the possibility of extraction of $R_{cr}$ value from the observed size dependence of a property. Keeping in mind that every $\overline{R}$ corresponds to one sample with its own distribution function, defined by $\sigma$ and $R_0$ values, it is obvious, that the curves depicted Fig. 2 with fixed values of $\sigma$ for every curve were made for illustration of the sizes distribution role rather than for description of any real experiment. On the other hand we have shown by fitting the observed in [11] $\varepsilon(\overline{R})$ dependence for BaTiO$_3$ nanoceramics by Eqs. (12a), (8a) that the condition (15a) is not satisfied for the majority of the $\overline{R}$ point (but two points in the "tails" of the curve) and the condition (15b) is not satisfied for all $\overline{R}$ experimental values (see Table 2 and Fig. 6), keeping in mind that for BaTiO$_3$ $R_{cr}(T = 300$ K$) = 16$ nm. Note, when calculating the solid curve in Fig. 6 we used $\delta = 0.001$ in Eq. (12a) and took the maxiaml intensity as a fitting parameter, that lead to pretty good description of all $\varepsilon(\overline{R})$ values.

Table 2. The values of parameter $\sigma$ extracted from experimental data for $\varepsilon(\overline{R})$ of BaTiO$_3$ nanograin ceramics [11]

| $\overline{R}$, nm | 350 | 450 | 550 | 650 | 750 | 1000 | 1250 | 1500 | 2250 | 3250 | 4250 | 5500 |
|---|---|---|---|---|---|---|---|---|---|---|---|---|
| $\sigma$, nm | < 100 | 150 | 350 | 550 | 850 | 1650 | 2000 | 1750 | 1750 | 2300 | > 2000 | > 2300 |
| $R_0$, nm | 350 | 450 | 541 | 599 | 562 | 159 | 324 | 1082 | 2126 | 3146 | 4250 | 5500 |

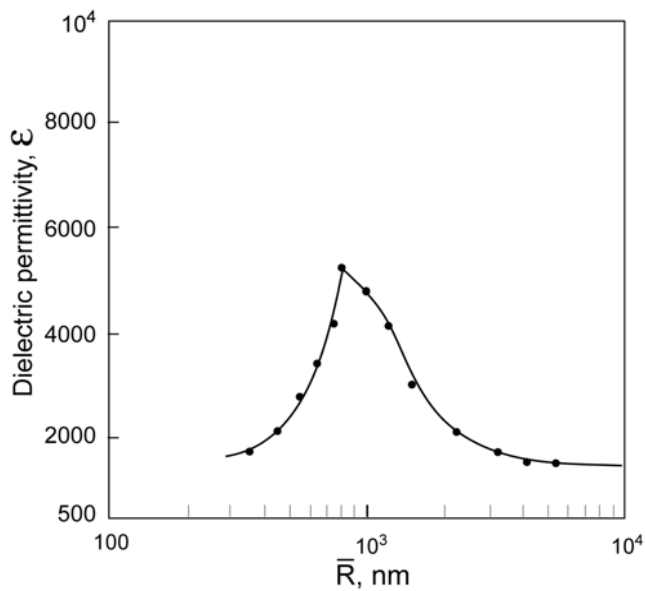

Figure 6. Size dependence of dielectric permittivity of BaTiO$_3$ nanograin ceramics. Solid line – theory, points – experiment [11].

In general case the condition (15b) is more hard, than (15a). Therefore the obtained essential dependence of dielectric permittivity maximum position on particles sizes distribution function width $\sigma$ and the estimation of the conditions (15) can pour light on the "puzzle" of much smaller (about 10 times) $R_{cr}$ value in nanopowder than in nanograin BaTiO$_3$ ceramics, derived from observed $\varepsilon(\overline{R})$ maxima position [11]. It should be stressed, that the conditions (15a) and (15b) are satisfied for the ceramics used for specific heat measurements as one can see from Table 1.

In temperature dependence $\varepsilon(T)$ the shift of $T_{max}$ to larger temperature with $\sigma$ increase was obtained also (see Fig. 3). While the decrease of the value of maxima $\varepsilon(\overline{R} = \overline{R}_{max})$ with $\sigma$ increase was obtained for all the considered $\sigma$ values (see Fig. 2) in $\varepsilon(T = T_{max})$ values there were both the decrease (see curves (1–4) in Fig. 3) and increase (see curves 5, 6 in Fig. 3) with $\sigma$ increase. The latter "peculiarities" is related to the case $T_{cl} \to T_c = 393$ K as one can see from Eq. (11a), because larger $\sigma$ value corresponds to bulk material. Because of the distribution of $T_{cl}$ it seems to be possible to extract from experimental data the most probable transition temperature $T_{cl}^0 = T_{cl\,max}$ only. This was confirmed by specific heat measurements. Indeed, the $T_m = T_{cl\,max}$ values obtained from specific heat maxima positions (see Table 1) were fitted pretty good by Eq. (3) at $R = R_0 = \overline{R}_{max}$ (see Fig. 4). From Fig. 3 for temperature dependence of dielectric permittivity one can see, that the parameters $R_0/\sigma$ for the curves 1, 2, 3 satisfy the condition (15a), while the others do not satisfy it. One can see, that $T_m$ for the curves 1–3 is close to one another and to the value of $T_{cl}^0$, i.e. to the most probable transition temperature. This is similar to the specific heat case. Keeping in mind, that the different curves in Fig. 3 correspond to different samples with different $\overline{R}$, the smearing of the curves 3, 4, which correspond to smaller $\overline{R}$, looks like the behaviour of specific heat also.

From general point of view the essential influence of sizes distribution function characteristics $\sigma$ and $R_0$ on dielectric susceptibility maximum position and height as well as on specific heat open the way for extraction of $R_0$ and $\sigma$ from experimental data, as it was shown in section 4. These parameters are very important for description of the properties in real nanomaterials and critical parameters of size driven phase transition.